\documentclass[]{spie}  

\usepackage{pbox}
\usepackage{multirow}
\usepackage{amsmath,amsfonts,amssymb}
\usepackage{graphicx}
\usepackage{floatrow}
\usepackage[export]{adjustbox}
\graphicspath{{.}} 
\usepackage{subcaption}
\usepackage{sidecap}
\usepackage[colorlinks=true, allcolors=blue]{hyperref}
\usepackage{enumerate}
\usepackage[flushleft]{threeparttable}
     
\title{Panoramic optical and near-infrared SETI instrument: overall specifications and science program} 

\author{
Shelley A. Wright\supit{a,b}, Paul Horowitz\supit{c}, J\'er\^ome Maire\supit{a}, Dan Werthimer\supit{d,e}, Franklin Antonio\supit{f}, Michael Aronson\supit{g}, Sam Chaim-Weismann\supit{d}, Maren Cosens\supit{a,b}, Frank D. Drake\supit{h},  Andrew W. Howard\supit{i}, Geoffrey W. Marcy\supit{d}, Rick Raffanti\supit{j}, Andrew P. V. Siemion\supit{d,h,k,l}, Remington P. S. Stone\supit{m}, Richard R. Treffers\supit{n}, Avinash Uttamchandani\supit{o} 
\skiplinehalf
\supit{a} Center for Astrophysics \& Space Sciences, University of California San Diego, USA; \\
\supit{b} Department of Physics, University of California San Diego, USA; \\
\supit{c} Department of Physics, Harvard University, USA; \\
\supit{d} Department of Astronomy, University of California Berkeley, CA, USA; \\
\supit{e} Space Sciences Laboratory, University of California Berkeley, CA, USA; \\
\supit{f} Qualcomm Research Center, 5775 Morehouse Dr, San Diego, CA, USA; \\
\supit{g} Electronic Packaging Man, Encinitas, CA, USA \\
\supit{h} SETI Institute, Mountain View, USA; \\
\supit{i} Astronomy Department, California Institute of Technology, USA; \\
\supit{j} Techne Instruments, Berkeley, CA, USA; \\
\supit{k} Radboud University, Nijmegen , Netherlands; \\
\supit{l} Institute of Space Sciences and Astronomy, University of Malta;\\
\supit{m} University of California Observatories, Lick Observatory, USA; \\
\supit{n} Starman Systems, Alamo, USA; \\
\supit{o}Nonholonomy, LLC, Cambridge, USA
}

\authorinfo{Further author information: (Send correspondence to S.A.W.)\\S.A.W..: E-mail: saw@physics.ucsd.edu}

\begin{document} 
\maketitle 

\begin{abstract}
We present overall specifications and science goals for a new optical and near-infrared (350 - 1650 nm) instrument designed to greatly enlarge the current Search for Extraterrestrial Intelligence (SETI) phase space. The Pulsed All-sky Near-infrared Optical SETI (PANOSETI) observatory will be a dedicated SETI facility that aims to increase sky area searched, wavelengths covered, number of stellar systems observed, and  duration of time monitored. This observatory will offer an ``all-observable-sky" optical and wide-field near-infrared pulsed technosignature and astrophysical transient search that is capable of surveying the entire northern hemisphere. The final implemented experiment will search for transient pulsed signals occurring between nanosecond to second time scales. The optical component will cover a solid angle 2.5 million times larger than current SETI targeted searches, while also increasing dwell time per source by a factor of 10,000. The PANOSETI instrument will be the first near-infrared wide-field SETI program ever conducted. The rapid technological advance of  fast-response optical and near-infrared detector arrays (i.e., Multi-Pixel Photon Counting; MPPC) make this program now feasible. The PANOSETI instrument design uses innovative domes that house 100 Fresnel lenses, which will search concurrently over 8,000 square degrees for transient signals (see Maire et al.\ and Cosens et al., this conference). In this paper, we describe the overall instrumental specifications and science objectives for PANOSETI.
\end{abstract}


\section{INTRODUCTION}\label{sec:intro}

\subsection{Background \& Motivation}\label{subsec:background}

Optical and infrared communication over interstellar distances is both practical and efficient. Just a year after the invention of the laser, it was suggested that laser technology could be used for optical communication over modest interstellar distances \cite{townes61}. Two decades later a detailed comparison of interstellar communication at a range of electromagnetic frequencies was explored\cite{townes83}, showing that optical and infrared wavelengths were just as plausible as the usual microwave/radio frequencies favored by SETI (Search for Extraterrestrial Intelligence) strategies of that era \cite{Cocc59}.

Lasers and photonic communication have improved considerably since then, with continuous wave laser power reaching into the megawatt regime, and pulsed laser power up to petawatts. Both continuous wave (CW) and pulsed lasers are plausible candidates for technosignature searches. CW and high duty cycle laser pulses could be easily detected with high-resolution spectroscopic programs that target individual stars \cite{Tellis2017}. Collimated with a Keck-size telescope, pulsed laser signals can also be detected: the pulses maybe can be orders of magnitude
brighter than the entire broadband visible stellar background \cite{Howard2004}. As an example, if we consider an ETI that transmits a 1PW laser with a 1~ns pulse width every ${\sim}10^4$ seconds to a set of target stars, a receiving civilization conducting an all-sky search would see the flash $\sim$10$^4$ times brighter than its host star. In this scenario, the sending civilization expends only 100W average power per target. The basis of this capability has already framed optical SETI search parameters for over two decades. 

Using current technology, pulsed optical SETI searches have the flexibility of being either targeted or covering large areas of the sky. An optical \textit{targeted} search of integrated visible spectra using the Automated Planet Finder (APF) at Lick Observatory is highly sensitive to continuous wave (CW) lasers and high duty cycle pulses ($>$ 1 Hz) from individual stars\cite{Tellis2017}. Spectroscopy is limited to targeted searches, with little possibility of a large field of view survey \textit{or} an all-time SETI search. Combining both pulsed and CW SETI targeted searches, the community has surveyed $>$10,000 stars with no detection \cite{Horowitz2001, Werthimer2001, Covault2001, Wright2001, Reines2002, Howard2004, Stone2005, Howard2007, Hanna2009, Wright2014, Abe2016, Sch2016}, although the dwell time per source observed has been very low ($\sim$10 min). Targeted SETI is poorly matched to intermittent signals sent by ET, and neglects millions of nearby stars that fall outside of the typical SETI target lists, as well as other potential astrophysical sources. There has been one wide-field optical (350 - 800 nm) SETI program that used a dedicated telescope for scanning the sky, but this search also had low dwell times \cite{Horowitz2001}. The missing link for laser SETI searches is the capability of continuous observations with large sky coverage, to increase phase space searched and likelihood of detection.

Extending the search into the near-infrared offers a unique window with less interstellar extinction and less background from our galaxy than optical wavelengths, meaning signals can be efficiently transmitted over larger distances. The infrared regime was specifically identified as an optimal spectral region for interstellar communication \cite{townes83}, yet has remained largely unexplored territory for SETI. The challenge has been lack of adequate near-infrared fast response ($\sim$ ns) sensitive detectors. Infrared detector technology has matured rapidly in the last decade, offering higher quantum efficiency and lower detector noise. Taking advantage of recent progress with infrared detectors, we developed the first near-infrared (950 to 1650 nm) SETI experiment that made use of the latest avalanche photodiodes for a targeted pulsed search \cite{Wright2014,Maire2014,Maire2016}. This program has motivated our team to develop both wide-field optical and near-infrared SETI instrumentation.

\begin{figure}[htb]
\includegraphics[width=0.8\textwidth]{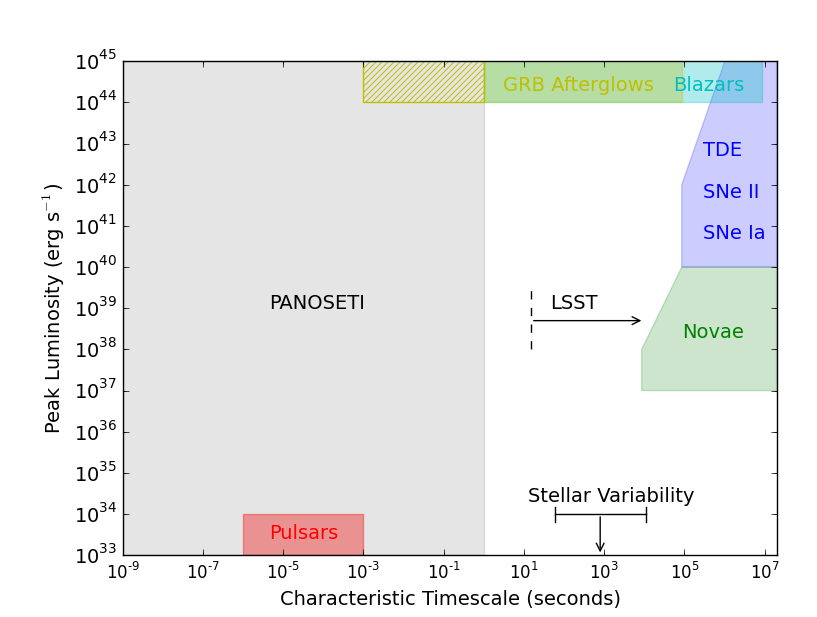}
\caption{Time domain of \underline{optical} astrophysical transients and variable sources: pulsars, supernova (Type Ia, II) and Tidal Disruption Events (TDE), classical novae, gamma ray burst afterglows, Blazars, and stellar sources. Fast time resolutions (nano-seconds $-$ seconds) have barely been explored and represents an observational limit with current ground and space-based facilities, especially since facilities are unable to represent large sky coverage with high duty cycles. Even with these limitations, new flavored transient sources are being found at shorter timescales (seconds), e.g., ASASSN-15lh. \cite{Cenko2017}. GRB afterglows can be observed for seconds to hours after the initial triggering event, but there have been no known observations that extend down to milli-seconds to seconds for rapid follow-up (hatched area). GRB 080319B \cite{Racusin2008}, the brightest recorded GRB in 2008, resides above the y-axis at $\sim$10$^{51}$ erg s$^{-1}$. Stellar variability from cataclysmic variables, Cepheids, stellar flares are typically $<$ 10$^{34}$  erg s$^{-1}$. The Large Synoptic Survey Telescope (LSST) will have unprecedented sensitivity, but its fastest time cadence is 15 seconds. PANOSETI will be capable of exploring luminous transient and variable phenomena from nanoseconds to seconds (grey area).
}\label{fig:transients}
\end{figure}

Wide field optical and infrared surveys make use of fast optics (i.e, low f/\#) apertures that are challenging to fabricate with good optical performance. Fast optical telescopes are expensive and need to make use of optical corrector lenses to reduce aberrations. Large optical surveys like Pan-STARRS \cite{Huber2015}, Zwicky Transient Factory \cite{Smith2014}, and the future Large Synoptic Survey Telescope \cite{Ivezic2008} have been designed to meet increasing interest in astrophysical transients and variable (repeating or stochastic) sources. In the last decade, new flavors of Type Ia and II supernova and novae have been discovered with optical transient surveys, expanding both observed lumonisities and characteristic time scales (see Figure \ref{fig:transients}). Typical instantaneous fields of view of these surveys are $\sim$ 10 - 50 sq.degrees, with a minimum time resolution of seconds-to-minutes. Gamma ray bursts (GRB) that trigger with space-based telescopes (e.g., Swift and Fermi) take minutes to hours for optical telescopes to respond for rapid follow-up of their afterglow. The fastest follow-up occurred on GRB 080319B (z=0.937), where data were taken within a few seconds because a wide-field optical imaging camera was already taking observations at the same sky location \cite{Racusin2008}. GRB 080319B was also one of the most energetic GRB events discovered with a peak visual magnitude of V=5.3 mag, making it even visible to the human eye \cite{Bloom2009}. The luminosity function of GRBs are still highly uncertain. The majority of optical counterparts have a typical peak magnitude range of V=10-18 mag within $<$ 1000 sec of follow-up \cite{Wang2013}.

At optical and infrared wavelengths, fast time (ms - $\mu$s) domain studies have been limited to targeted searches of already known variable sources like pulsars, cataclysmic variables, and extremely luminous stars. Extending to nanoseconds has been limited to a few sources like the Crab Pulsar \cite{Eikenberry1997, Leung2018}. Fast $>$GHz photometers are now being explored for quantum phenomena on future Giant Segmented Mirror Telescopes ($>$20m) where the aperture is sufficient to not be photon starved \cite{Barb2007,Shearer2008}. All of these current programs have a low duty cycle on-sky and will only make a few observations per field over the course of their operation. Many order of magnitudes in time scales are not currently covered by space- and ground-based optical observatories, as seen in Figure \ref{fig:transients}.

In contrast, radio observatories have dominated searches for fast transient and variable sources at milli- to micro-second time scales. Historically, radio transient observations have targeted single compact objects like pulsars, X-ray binaries, and active galactic nuclei. But with the recent discovery of Fast Radio Bursts (FRBs) \cite{Lorimer2007}, fast radio transient searches have enjoyed a boom. Discovery of the original FRB was made possible by re-processing archival data from the Parkes radio telescope and searching for transients at millisecond time scales. Once the time domain and luminosity of FRBs were known, subsequent discoveries easily followed using other wide-field radio telescopes \cite{Petroff2016}. Even though FRBs eluded discovery for decades, remarkably the implied rate is 10,000 events per day per $4\pi$ steradians (or 1 FRB per day per 4 sq.$\space$degrees) \cite{Thornton2013}. Despite many days devoted to covering large duty cycles of time on previously discovered FRBs, only once source, FRB 121102, has been found to repeat with multiple radio pulses\cite{Spitler2014}.  

With ground-based gravitational wave detectors in full operation, LIGO-Virgo \cite{Abbott2009} will be capable of discovering mergers of black hole binaries, neutron binaries, and black hole - neutron star binaries at distances of several Mpc. The possibility of electromagnetic follow-up is a prime directive of the time domain astronomy community. This has ignited the Multimessenger\cite{Smith2013} community that has developed multi-wavelength facilities on ground- and space-based observatories for rapid follow-up. For instance, in 2017 the Fermi satellite discovered GRB 170817A, and LIGO confirmed detection of a binary compact merger associated with the GRB\cite{Goldstein2017}. This was the first electromagnetic counterpart discovered of a gravitational wave event. An electromagnetic counterpart may be either precursor or an afterglow of the gravitational wave event, possibly a flash triggered during the merger event or the ring-down after the merger. Timescales and luminosities of such an electromagnetic counterpart event are largely unexplored, and current multi-wavelength surveys are only planning to achieve follow-up within minutes-to-hours from a gravitational wave event, with an observational time resolution of a few seconds. 

Optical and infrared SETI instrumentation that explores the very fast time domain, especially with large sky coverage, has a prime opportunity for new discoveries that complement Multimessenger and time domain astrophysics. In this paper, we describe design and plans for a new observatory network that is capable of searching for extremely rapid (nanosecond to second) optical and near-infrared events from either artificial or natural phenomena, over the entire ``observable" sky. This new observatory is being designed for a large-scale SETI experiment, and given its wide sky coverage and long duty cycles, it is equally capable of making new astrophysical discoveries within the fast time domain. 

\subsection{Program Objectives}\label{subsec:objectives}

This program aims at developing an ``all-sky" optical and wide-field near-infrared pulsed SETI experiment that is capable of surveying the entire northern hemisphere. The observatory design may easily be replicated for southern skies as well. The requirements for this program address \textit{seven} essential ``missing corners'' of current optical/infrared astrophysical transient and technosignature programs.

\begin{enumerate}
  \item Extending wide-field searches to the desirable near-infrared, boosting wavelength coverage by 1.7 octaves.
  \item Investigating the entire ``observable" sky, increasing instantaneous field coverage by a factor of 25,000 (Harvard \cite{Horowitz2001}) and 2,500,000 (Automated Planet Finder \cite{Tellis2015}).
  \item Adding first capability of an ``all-time" optical search, increasing the fraction of time observed on \textit{any} given source by a factor of 100,000.
  \item Enlarging the number of observed stellar sources to 100's of millions stars.
  \item Implementing search methods for pulse transients and variable sources over 10 decades: nanosecond to seconds.
  \item Operating the first dedicated, simultaneous all-sky, all-time dual optical SETI facility.  This duality is essential for unambiguous and immediate confirmation of a candidate signal (e.g., compare with gravitational wave detection with LIGO\cite{Abbott2009}).
  \item Explore a new time domain that is capable of revealing unknown astrophysical optical transient or variable phenomena arising from compact objects and mergers over nanosecond to second time scales.
\end{enumerate}
 
Herein, we describe overall specifications of the Panoramic optical and near-infrared SETI (PANOSETI) observatory and instrument system, which will achieve all of these objectives.  The preliminary design of the geodesic dome and Fresnel lens characterization are described in Maire et al. \cite{Maire2018}, and the opto-mechanical design of a single telescope aperture is presented by Cosens et al.\cite{Cosens2018}, this conference. 

\section{OBSERVATORY \& INSTRUMENT DESIGN}\label{sec:observatory}  

\begin{figure}[htb]
\centerline{\includegraphics[width=0.8\textwidth]{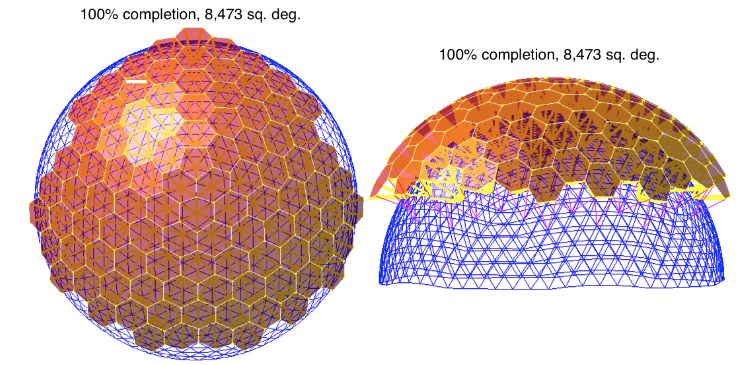}}
\caption{Conceptual design of a tessellation 12 frequency geodesic dome seen from above (left) and the side (right). This geodesic dome design is made of a tessellated shell that holds 100 Fresnel lens modules. The outer layer (red) are where Fresnel lenses are housed, and the inner-layer (blue) is where the detectors are placed. The geodesic dome allows modular addition of new Fresnel and detector units as they are completed during different stages of fabrication.}
\label{fig:dome}
\end{figure}

Our observatory design reduces the cost per aperture on sky while maintaining sky coverage and flexibility of wavelength bandpass. Instead of using a traditional optical telescope we have designed a system that uses Fresnel lenses (f/1) for each collecting aperture. We plan to construct a geodesic dome populated with $\sim$100 Fresnel modules with fast-response detectors \cite{Maire2018}. A geometric layout of a single geodesic dome is shown in Figure \ref{fig:dome} that highlights our potential sky coverage. 

\begin{figure}[htb]
\includegraphics[width=0.45\textwidth]{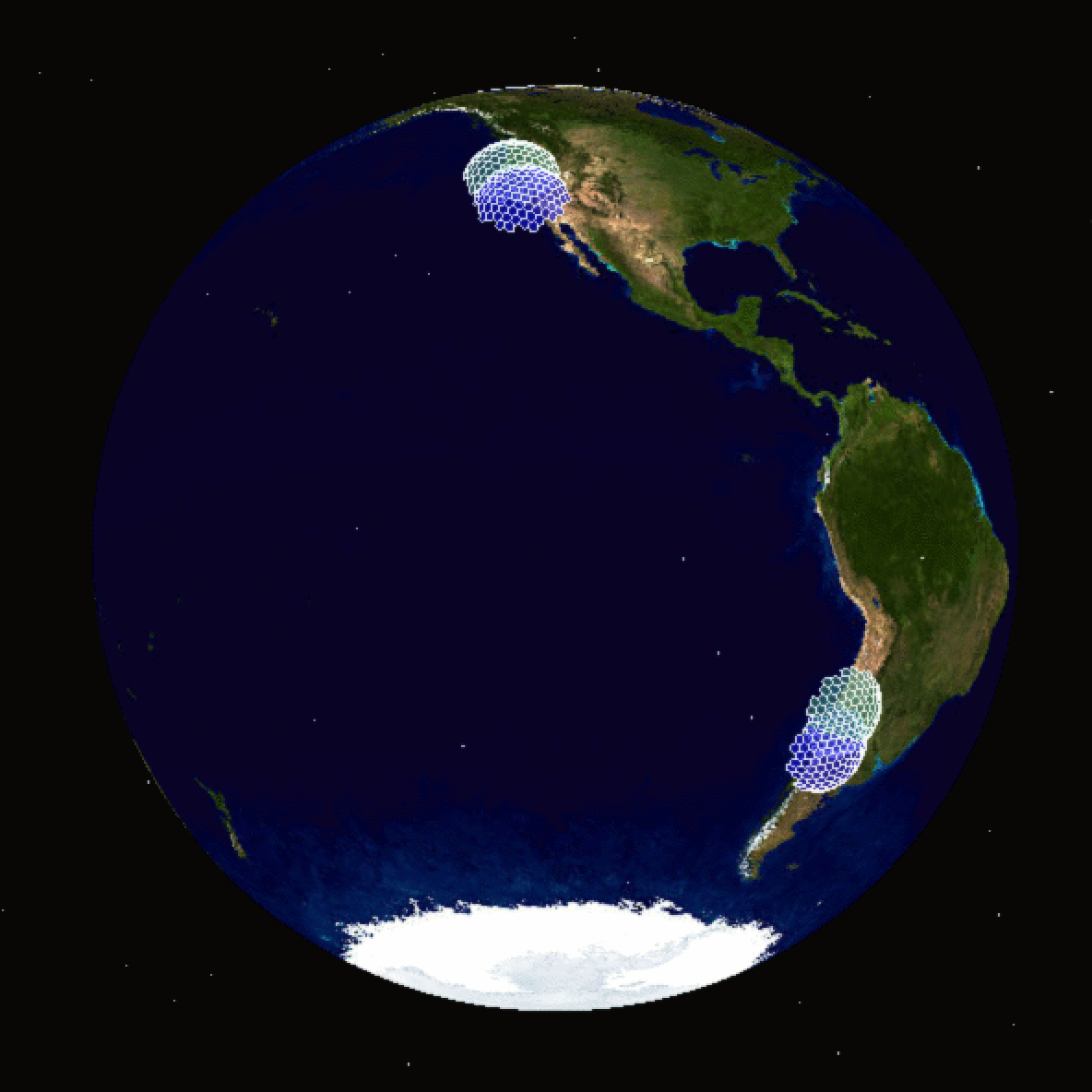}
\includegraphics[width=0.45\textwidth]{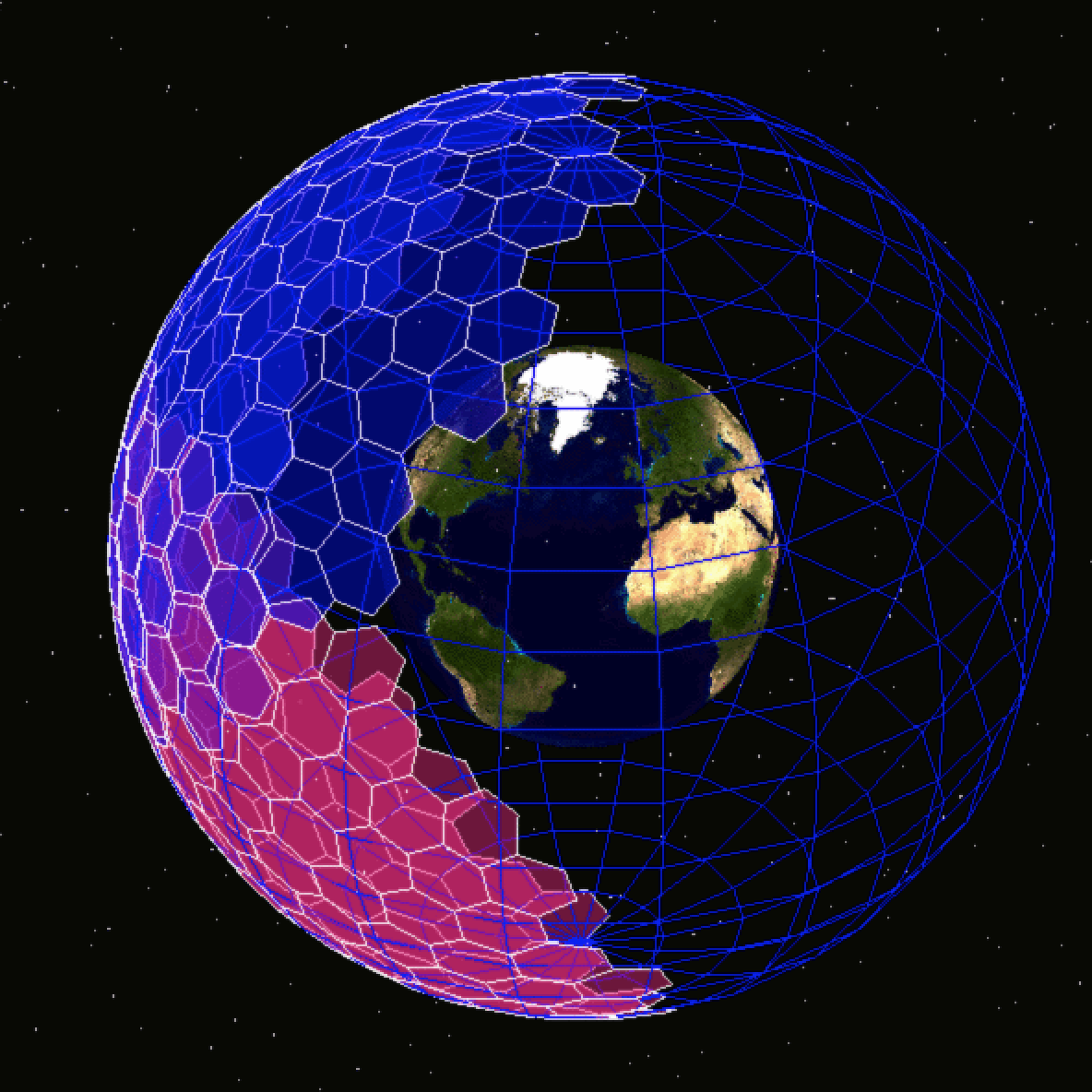}
\caption{(LEFT) A pair of optical and near-infrared PANOSETI dome facilities located in both the northern and southern hemispheres. Each facility is separated locally by $\sim$675 km and will be observing the same sky regions for coincidence detection. (RIGHT) Instantaneous field of view from both PANOSETI facilities projected on the celestial sphere. Each hexagon maps to a 0.5m Fresnel lens. The entire ``observable" hemisphere will be observed from both locations at optical wavelengths. Several lenses per site will be dedicated to near-infrared drift scan observations.}
\label{fig:earth}
\end{figure}

Our planned operation is to have at least two geodesic domes per hemisphere to allow for unambiguous detection of a source simultaneously visible from both sites, as shown in Figure \ref{fig:earth}. We believe this is imperative to rule out false alarms at a single observing site that would otherwise be affected by noise emanating from cosmic ray showers, atmospheric phenomena, or other site specific noise. Multiple copies of this instrument could be placed at more sites for both efficiency, hemispheric coverage, and confirmation of detection, as illustrated in Figure \ref{fig:earth}. A dome shelter will be dedicated to protection of each geodesic dome, and will operate autonomously with a dedicated weather center. 

\begin{figure}[htb]
\includegraphics[width=0.8\textwidth]{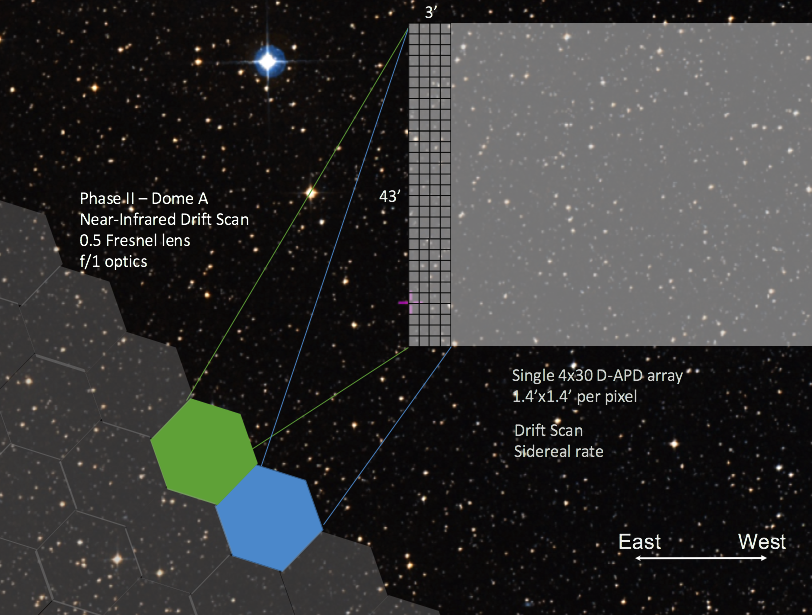}
\caption{ Two 0.5\,m lenses installed in the geodesic dome will be dedicated to a near-infrared wide field search. Each near-infrared module will use a custom-made 4$\times$30 near-infrared APD array with an individual pixel size of 200 $\mu$m. These modules would operate in a drift scan mode covering different elevations over time, and would be able to sample the entire northern hemisphere in 230 clear nights.}\label{fig:nirfov}
\end{figure}

The current design for optical wavelengths has each Fresnel module achieving a field of view of 9$^\circ$ $\times$ 9$^\circ$ with 20 arcminute per pixel, so the total geodesic dome will achieve an instantaneous sky coverage of $>$8,500 square degrees. In each geodesic dome, there will be at least two modules dedicated for a near-infrared wavelength wide-field search. Each near-infrared module will achieve 82.5 arcsecond per pixel with a total instantaneous field of view of 0.06 square degree. The near-infrared component will operate in drift scan mode, as seen in Figure \ref{fig:nirfov}. The geodesic dome allows us to move the near-infrared modules in elevation along the meridian and is capable of mapping the entire observable sky in 230 clear nights.

We have structured our instrument development to be conducted in multiple phases that aim to reduce risk and complexity of the project. We are currently in the preliminary design phase and are developing an end-to-end prototype of the opto-mechanical Fresnel module and detector electronics, as well as finalizing major design for the observatory. Instrument specifications for both the optical and near-infrared components are summarized in Table \ref{tab:param}.

\begin{table}[htb]
\caption{Specifications of the SETI All-Sky  Optical \& Near-Infrared Experiment. Number of components are given for \textbf{one} site.}
\label{tab:param}
\begin{center} 
\begin{tabular}{|l|l|l|l|l|}
\hline 
\multirow{2}{*}{\textbf{PARAMETER}} &  \multicolumn{4}{c|}{\textbf{VALUES}} \\
&  \multicolumn{2}{c|}{\textbf{Visible}} & \multicolumn{2}{c|}{\textbf{Near-Infrared}}\\
 \hline 
  Light collecting areas & \multicolumn{4}{c|}{$\sim$100 x 0.5-m f/1 Fresnel lenses}   \\
   \hline
  Detectors & \multicolumn{2}{l|}{\pbox{6cm}{Hamamatsu arrays 3mm pixels\\
   (MPPC-S13360-3025CS)  }} & \multicolumn{2}{l|}{\pbox{6cm}{Amplification Tech.\\
   InGaAs 200$\mu$m-pixels 30x4 pixel array  } } \\
 \hline
  Sensitivity  &   \multicolumn{2}{l|}{25 photons per pulse} & \multicolumn{2}{c|}{100 photon/pulse (5\% false alarm rate/night)} \\
     \hline
  Wavelength coverage  &   \multicolumn{2}{l|}{300 $-$ 850} & \multicolumn{2}{c|}{850 $-$ 1650} \\
  \hline
  Time waveform resolution & \multicolumn{4}{c|}{$\sim$1\,ns, rise and fall} \\
  \hline
   Target Goal  &   \multicolumn{2}{l|}{Observable sky} & \multicolumn{2}{c|}{Drift scan mode} \\
  \hline
  Plate scale &   \multicolumn{2}{l|}{0.36 degree per 3mm pixel }& \multicolumn{2}{l|}{ 82.5 arcsec/pixel (6.8 arcmin/mm) }\\
 \hline
    Sky coverage & 
    \multicolumn{2}{c|}{
  $>$ 8,000 sq.\ deg.\
}
  &
   \multicolumn{2}{|l|}{\pbox{10cm}{0.06 sq.\ deg.\ in drift scan mode,\\
   all-(observable) sky in  230 clear nights  }
 }\\
 \hline
   \textit{Minimum} Dwell time  &  \multicolumn{2}{c|}{ All-(observable) time} & \multicolumn{2}{l|}{ 22s  (all detectors), 5.5s (pixel)}  \\
 \hline
\end{tabular}
\end{center}
\end{table}

\subsection{INDIVIDUAL MODULE}\label{sec:modules}

\begin{figure}[htb]
\centerline{\includegraphics[scale=0.35]{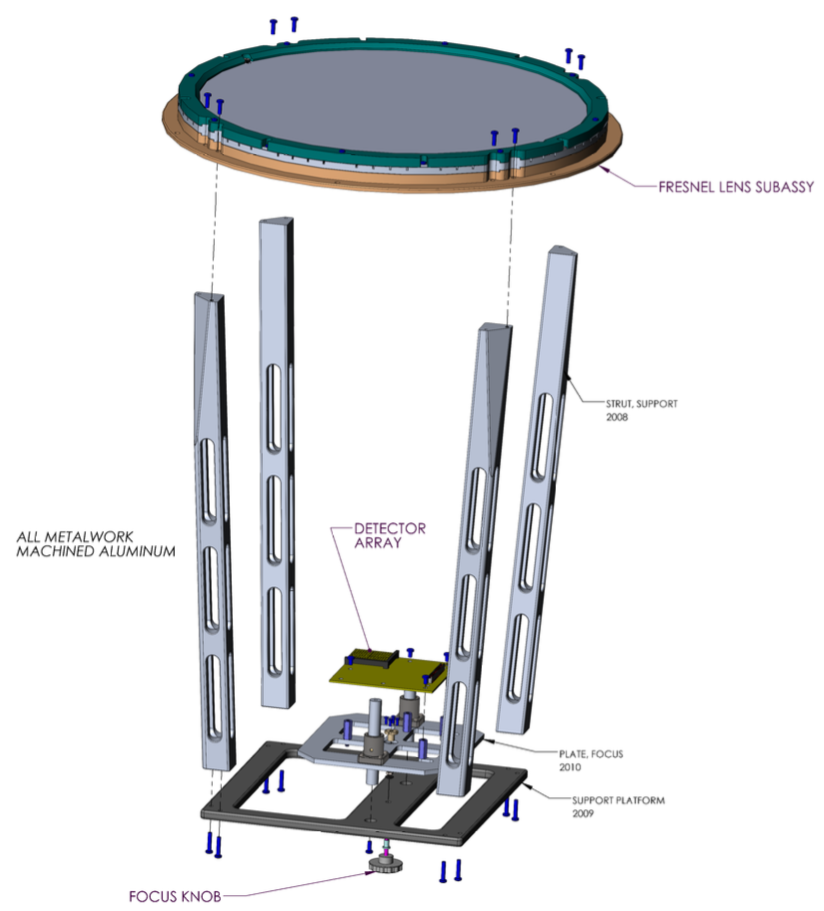}}
\caption{An individual module with Fresnel lens mount (0.5m), struts, and optical 32x32 Hamamatsu MPPC optical detectors that are being designed for the prototype unit. The focus stage is manual during this preliminary design phase as we test the equipment on-site.
}\label{fig:module}
\end{figure}

Each Fresnel lens will be housed in a module unit (Figure \ref{fig:module}) that will baffle stray light. We define a module as an aperture unit that contains both the Fresnel lens and detector plane. We have now designed a prototype module that includes opto-mechanical housing of the 0.5m Fresnel lens, detector mount, focus stage, and position angle mechanism. 

The opto-mechanical design and thermal analysis of the individual module is presented in Cosens et al.\cite{Cosens2018}. Briefly, the opto-mechanical lens mount includes a protective acrylic transparent cover to protect the Fresnel lens from dust and scratches. The air gap between Fresnel lens and protective cover reduces potential condensation. Four struts connect the lens mount to the detector housing and back-end electronics. A focus stage at the detector plane derives initial focus and compensates for focus shifts at the telescope due to temperature changes or drift. Current design of the prototype unit is shown in Figure \ref{fig:module}.

Both on-axis and off-axis spot sizes of the Fresnel lenses are well-matched to the optical pixel sizes and small shifts in focus will not impact the optical performance. Chromatic aberration dominates the aberration terms for the Fresnel lens. We will determine focus at the central wavelength for the optical and near-infrared bandpasses. We further describe the optical quality and characterization of Fresnel lenses at optical wavelengths in Maire et al. \cite{Maire2018}. Given the smaller near-infrared pixel sizes (200 $\mu$m) we are exploring a corrector lens for these Fresnel modules.

\subsection{DETECTORS}\label{sec:detectors}   

We will use multi-pixel photon counter (MPPC) detectors for optical (300--850\,nm) and near-infrared (850--1650\,nm) wavelengths. An MPPC is an array of independent Geiger-mode avalanche photodiodes (APD), whose outputs are summed to a single terminal; this single pixel exhibits excellent pulse-height resolution, since each sub-pixel generates either a fully saturated pulse or is dormant.

\subsubsection{\textbf{Optical detector arrays (300 –- 850 nm)}}\label{subsec:opticaldetector}

For the optical we are using the Hamamatsu silicon photomultipliers (SiPM), which are photon-counting devices using multiple APD pixels. Each Fresnel lens will illuminate a focal plane tessellated with 32x32 optical MPPCs in an array covering an instantaneous field of view of 81~deg$^2$. Four 8$\times$8 SiPM arrays with individual 3mm$\times$3mm pixels (i.e., Hamamatsu p/n S13360-3050CS) will be tiled 2$\times$2 on the optical axis, thereby achieving a continuous 32$\times$32 pixel field of view.

\begin{figure}[htb]
\centerline{\includegraphics{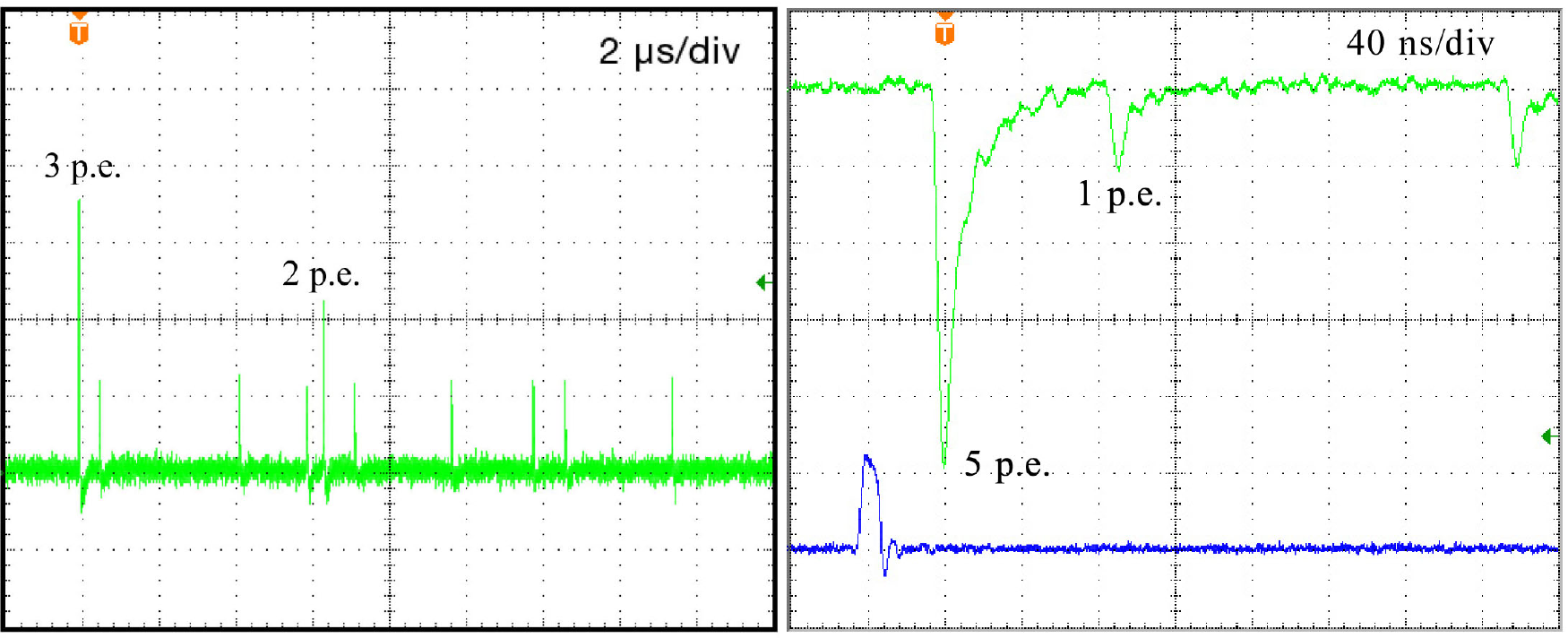}}
\caption{LEFT: Dark counts from an MPPC (Hamamatsu 3\,mm square, with 14,400 25\,$\mu$m cells) operated at recommended bias, loaded into 50\,$\Omega$ and amplified with a UTO-style RF amplifier. Note the clean discrimination of single photoelectrons (p.e.), excellent linearity, and low noise floor. The division is 2 $\mu$s with 3 and 2 p.e. events identified.
RIGHT: A test flash (lower trace) is detected consistently
(here at the 5 p.e. level, upper trace), even in the
presence of background illumination producing 10$^{6}$ counts per second rate or more of single-p.e.\ events. (The 40\,ns delay is caused by a lengthy fiber run from LED flasher to dark-box.)}
\label{fig:pulser_test}
\end{figure}

Our design with its pixels subtending 0.36$\times$0.36 degrees or 20x20~arcmin (thus 400 square arcmin) sees a sky brightness of $\sim$ $m_V{\approx}6.6$, thus about $1.4{\times}10^7$ visible photons per second, reduced by photon detection efficiency (PDE) to $\sim$ $4{\times}10^6$ single p.e.\ counts per second. We performed a series of bench-top experiments with a single 3mm$\times$3mm SiPMs to measure the dark current, background count rate and performance. The dark count rate of the optical MPPCs are of order $5\times 10^5$ counts per second. We have observed preliminary background count rates with a prototype 0.5m Fresnel lens and a single pixel at Mt. Laguna Observatory during different lunar conditions. Given the large field view per pixel, we will be background-limited from the night sky, stellar background count rate, and moon illumination. Moon illumination will be a dominant source of background noise in the experiment. Theoretical calculations for our projected sites (Mt. Laguna and Lick Observatory) match our observations on-sky. We found during no moon an integrated  background (dark + sky + stellar) of 12x10$^6$ counts per second, and during partial illumination of (50\% lunar illumination) we expect 4x10$^7$ counts per second (at over 10$^{\circ}$ away from moon). This is roughly 25 times the room-temperature dark count rate per pixel as measured in the laboratory (see Figure \ref{fig:pulser_test}) and manageable. 

\begin{figure}[htb]
\includegraphics[width=0.9\textwidth]{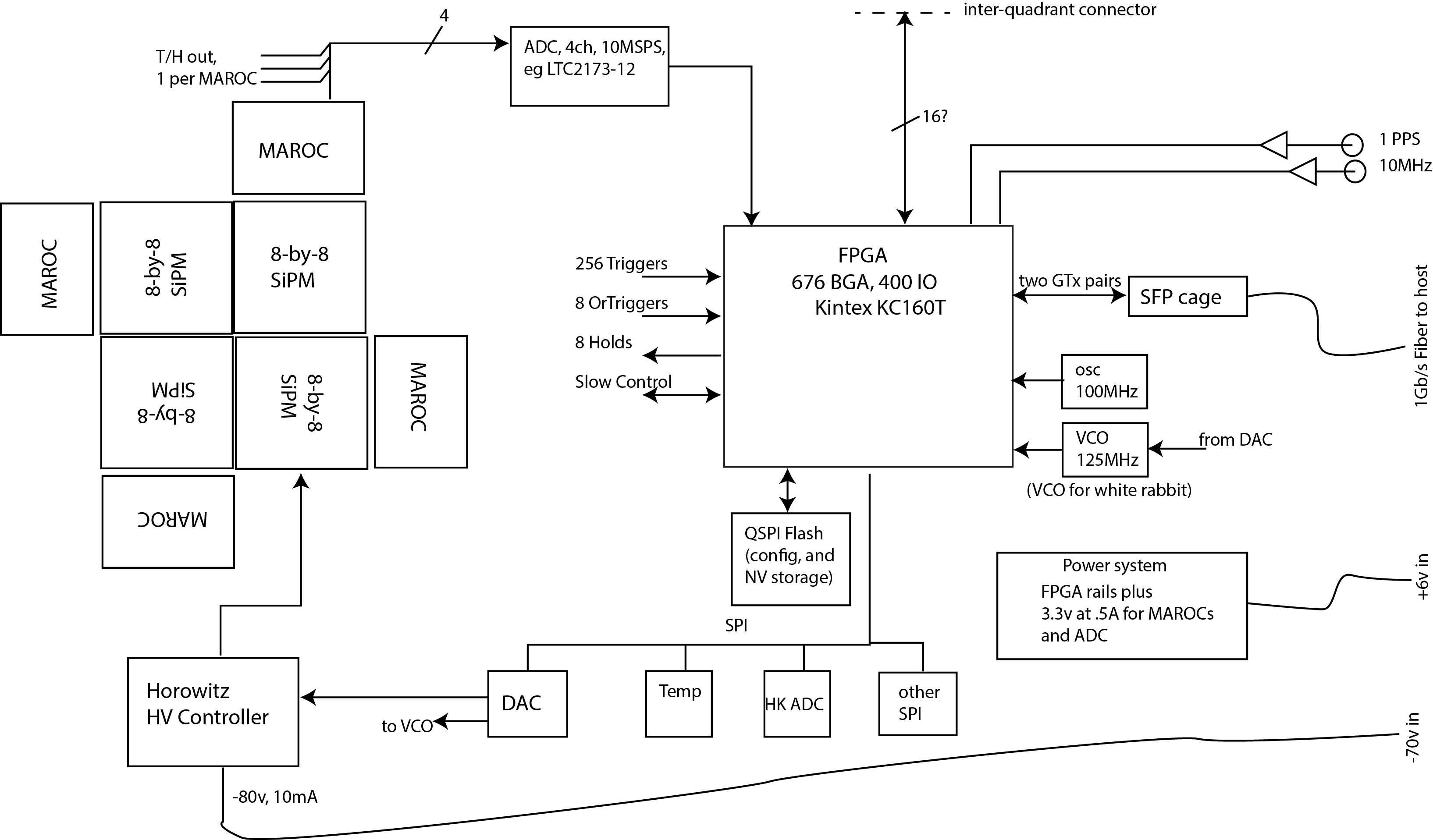}
\caption{Schematic of the prototype quadrant board to readout four optical Hamamatsu 8$\times$8 pixel SiPM detector arrays. The block diagram highlights the MAROC-3A ASICs, ADC, FPGA, custom-made high voltage power supply for the detectors, and SFP output to a 1Gb/s connection. We will make use of White Rabbit PTP that makes VCO (125 MHz) for high-precision time stamping to 1 ns in order for coincidence detection between the pairs of observatory.
}\label{fig:block}
\end{figure}

MPPC detectors are good at detecting single photon events even with the high background count rates we expect with varying sky conditions at optical wavelengths. We are designing a custom readout board that makes use of 64-bit Application-Specific Integrated Circuit (ASIC) Weeroc Maroc-3A that is capable of pulse shaping and providing trigger detection of individual pulses. As shown in Figure \ref{fig:block}, we have designed our prototype detector  board with four 8$\times$8 pixel SiPM arrays that each feed a MAROC-3A ASIC that amplifies the signal from the SiPMs and provide a per-pixel trigger signal to a Kintex Field Programmable Gate Array (FPGA). A four-channel high voltage controller provides precise high voltage bias to each SiPM array. A 10$^{6}$ sample per second analog-to-digital converter (ADC) will read the detected events of all pixels whenever any single pixel is triggered. A 1 Gb per second fiber connection provides data communication to a host. We are planning to use White Rabbit\cite{Moreira2009} precision time protocol  functionality that will provide $\sim$1ns time-stamping for each event. The four quadrant boards will be arranged to form a 32$\times$32 array field of view.

The proposed optical pulsed SETI makes no assumptions about particular laser technologies, being sensitive to all wavelengths in the detector's passband (400--1000\,nm for current Hamamatsu MPPCs). A civilization 500 light-years distant, with only our (primitive) level of technology, could launch nanosecond bursts of 25 visible photons into our 0.5\,m aperture, enough to trigger the system.  Then, as described in the scenario at the beginning of this proposal, this system would reliably detect such an ETI within that distance anywhere in the northern sky willing to devote 100W per target. 

\subsubsection{\textbf{Near-infrared Detector Arrays (850--1650\,nm)}}\label{subsec:nirseti}

There has never been a \textit{wide-field} near-infrared SETI program. To achieve this we need to work with the most available, advanced detectors that operate cryogenically. We plan to have at least four near-infrared arrays; two in each PANOSETI observatory. Our current detector source (Amplification Technologies, a division of Powersafe Technology Corp.) has made crucial advances in semiconductor detection technology for extremely high sensitivity, high bandwidth photon detection. Their near-infrared InGaAs detectors (850$-$1650 nm) have low noise, and are fast (1~GHz) with very low quench times ($<$1 ns). For final production we are designing around a custom-made array from Amplification Technologies that will have a single package thermal-electrically cooled (-30$^\circ$C) APD array with configuration of $30{\times}4$ pixels and pixel pitch of 200\,$\mu$m. During the prelimiary design phase our team is making use of 5x5-pixel APD arrays from Amplifcation Technologies for characterization and prototyping. 

Unlike the optical experiment, at nanosecond resolution the near-infrared arrays are dark-count limited rather than stellar photon-limited (for stars fainter than J=1 in a 1-m telescope configuration, according to NIROSETI measurements\cite{Wright2014}). Sensitivity benefits from relatively low dark-noise in the near infrared (less than 1x10$^{6}$ counts per second). Near-infrared sky background is dominated by OH emission lines generated by reactions between O$_3$ and H in the upper atmosphere. Strength of the OH emission is site-dependent and highly variable on minute timescales. Sky brightness due to the Moon also varies by 0.9 magnitude in J-band and 1.2 magnitude in H-band depending on the lunar cycle \cite{Vanzi2003}. Using the projected night sky brightness at Mt. Hamilton (elevation 1300 m) we find that the maximal field-of-view per pixel beyond which the sensitivity of the instrument is sky-background limited for a 0.4-m, 0.5-m, and 1-m aperture is respectively 175x175 arcsec, 140x140 arcsec and 70x70 arcsec. For the selected 0.5m Fresnel lens with our near-infrared detectors, the plate scale is 91 arcsec per $200\mu m$-pixel, which will satisfy the background requirements for this system at a dark site. 

\section{ANALYSIS METHODS}\label{sec:analysis} 

To detect optical pulse widths ranging from nanoseconds to seconds, the PANOSETI readout electronics system for each module will be configured to support three different operational modes.

\textbf{Pulse Height Mode} is optimized for detecting optical pulse widths $<$ 30~ns.  Pulse Height Mode takes advantage of the fact that stars are dim at nanosecond time scales, so a short bright optical pulse produces effective photon pileup during the resolving time of the detector/amplifier/shaper circuit, resulting in a pulse height that is easily discriminated against the single-photon events from the stellar background. Pulse height data acquisition occurs when any of the 256 triggers from the Maroc exceeds a programmable threshold. The pulse heights from all 256 peak-holds are digitized as well as all 256 baseline voltages. The data is time stamped, packetized and sent to the computer over ethernet, see Figure  \ref{fig:block}. The Pulse Height Mode has been standard for previous optical SETI experiments \cite{Wright2001, Werthimer2001, Horowitz2001, Howard2004, Wright2014}. It is a useful mode for detecting multiple photon events with pulse widths $<$ 30~ns, and can be used to compared directly with calibration pulse height distributions \cite{Maire2016}. This mode is well-suited for technosignature search and discovering new astrophysical phenomena.

\textbf{Continuous Imaging Mode} employs counters on every pixel that count over-threshold-events to produce images at a programmable frame rate. This mode will be used for detecting transients of 100 $\mu$s or longer, where the data and computing rates are slow enough that software can be used to search for transients. Over-threshold events for each detector are counted by the FPGA. Each 24-bit counter is capable of count rates up to 200x10$^6$ counts per second. At a programmable frame rate all pixel counters are sent by the FPGA over 1Gb ethernet to the central computer system, along with an accurate time stamp for each frame. Continuous Imaging Mode will also be used for calibration of the telescope pointing and detector pointing, by observing stellar positions as the earth rotates (e.g., measuring precisely when a bright star enters a pixel and when it exits). The continuous imaging mode is well matched to luminous astrophysical transients and variable sources, especially gamma ray bursts.

\textbf{Triggered Imaging Mode} is similar to Continuous Imaging Mode, but to keep the data rate and computing requirements manageable at high frame rates, each module's image is transmitted and stored only when one or more of the pixel counters is found to have an excess count rate. This mode will be utilized for detecting transients from 100 ns to 100 $\mu$s, where the frame rates are too high for software to search for pulsed events. At these high cadence frame rates, the FPGA gateware and embedded processor are utilized to search for transient events. The gateware software decides when a pixel counter statistically significantly exceeds the average count; then the board transmits the image frame for the event along with an accurate time stamp and the statistical parameters used to detect the event. This time domain has been largely unexplored by optical SETI experiments and its new capabilities will expand the phase space searched with the potential for unraveling other astrophysical phenomena. 

All individual ``events" will be stored in a central computer for each observatory. Individual events will then be cross-correlated in time and sky localization between the two observatory domes. A coincident event (time, right ascension and declination) between the two observatories would become a ``candidate". This candidate event can be matched with identified sources from the Multimessenger programs for any astrophysical transient event. In fact if just one significant event is flagged at a single observatory, we have the ability of correlating this with other transient events (e.g., Fermi Satellite, LIGO). For each candidate, we would plan follow-up with higher spatial resolution imaging observations for isolating a counterpart, and would trigger other SETI facilities to observe that sky region. Given our use of White Rabbit timing protocol, if we measure the time delay between the two observatory events, where the observatories are $\sim$700 km apart, then we can improve the astrometric localization of the source down to 2-3 arcseconds along one spatial axis.

\section{SUMMARY}\label{sec:summary}

Current astronomical wide-field sky surveys have poor sensitivity to optical transients or variable sources with a duration less than a second, as most sky surveys utilize low-noise (CCD or CMOS) cameras that integrate for several minutes or longer. This largely unexplored phase space of sub-second optical and near-infrared pulse widths is perfectly suited for a technosignature survey and will enable new discoveries of astrophysical transient and variable phenomena. 

We are developing a novel optical and near-infrared program that will dramatically improve the search for fast (nanosecond - seconds) transient or variable events, with large sky coverage, wavelength bandpass, and duty cycle per source. Our proposed experiment, deployed at paired observing sites (to eliminate ``false alarms''), would provide all-time coverage of a substantial portion ($>$8,500\,deg$^2$) of the observable night sky. At near-infrared wavelengths, we will perform the first ever wide-field SETI survey, extending to prime wavelengths for interstellar communication.

Our team is currently in the preliminary design phase of the program. The first phase will be dedicated to developing prototype systems, as well as finalizing all major designs for both optical and near-infrared modules and observatory structure. We have acquired Fresnel lenses and conducted laboratory tests demonstrating the necessary optical quality for the fast-response detectors, and have confirmed capabilities of both optical and near-infrared detectors. Our team will build at least four Fresnel modules in this phase to develop on-sky experience and to advance our software and post-processing techniques. We will deploy dual prototype modules at two sites (Mt. Laguna and Lick Observatory) to test our coincidence techniques. Our geodesic dome structure will allow modular deployment of individual apertures, and we anticipate that PANOSETI will have two dual sites commencing operation within 3 years.

\acknowledgments       
 
The PANOSETI research and instrumentation program is made possible by the enthusiastic support and interest of Franklin Antonio. We thank the Bloomfield Family Foundation for supporting SETI research at UC San Diego in the CASS Optical and Infrared Laboratory. Harvard SETI is supported by The Planetary Society. UC Berkeley's SETI efforts involved with PANOSETI are supported by NSF grant 1407804, the Breakthrough Prize Foundation, and the Marilyn and Watson Alberts SETI Chair fund. We would like to thank the staff at Mt. Laguna and Lick Observatory for their help with equipment testing. The following vendors have been incredibly gracious with their answers to our frequent technical questions: Hamamatsu, Weeroc, Caen Technologies, and Amplification Technologies. 
 
\bibliography{report}
\bibliographystyle{spiebib}

\end{document}